\documentclass[prb,aps,twocolumn,superscriptaddress,showpacs]{revtex4}
\usepackage{latexsym}
\usepackage{graphicx}
\usepackage{amsmath}

\begin{document}

\title{Influence of orbital nematic order on spin responses
in Fe-based superconductors}

\author{Yuehua Su}
\affiliation{ Department of Physics, Yantai University, Yantai 264005, P. R. China}

\author{Chao Zhang}
\affiliation{ Department of Physics, Yantai University, Yantai 264005, P. R. China}

\author{Tao Li}
\affiliation{ Department of Physics, Renmin University of China, Beijing 100872, P. R. China}

\begin{abstract}
Electronic nematicity is ubiquitous in Fe-based superconductors, but
what the primary nematic order is and how the various nematic phenomena
correlate with each other are still elusive.
In this manuscript we study the physical consequence of the orbital nematic order
on the spin correlations. We find that the orbital nematic order can drive a significant
spin nematicity and can enhance the integrated intensity of the spin fluctuations.
Our study shows that the orbital nematic order has strong effect on the spin correlations
and it can not be taken as an unimportant secondary effect of the nematic state
in Fe-based superconductors.
\end{abstract}

\maketitle

\section{Introduction} \label{sec1}

The electronic nematic state is a novel state in nature which breaks rotational
symmetry spontaneously but is translational invariant\cite{Fradkin}.
In the newly discovered Fe-based superconductors (FeSCs),
the rotational symmetry breaking is ubiquitous. It has been observed in different
experimental probes, charge resistivity\cite{ChuScience,ChuScience2012},
angle-resolved photoemission spectroscopy (ARPES)\cite{ZXShen,YiShen2011,Feng2012,YiShen2014,DingH2015,LuDH2015},
neutron scattering\cite{HarrigerDai2012,LuoDai2013,LuDaiScience2014,McQueeney2014},
optical conductivity\cite{Uchida2011,Lucarelli2011},
nuclear magnetic/quadrupole resonance (NMR/NQR)\cite{Fu2012,Baek2014,Lang2010},
magnetic torque measurement\cite{KasahasiMatsuda},
scanning tunneling microscope/spectroscopy (STM/STS)\cite{AllanDavis,Rosenthal,Singh2013},
Raman scattering\cite{Gallais2013},  X-ray diffraction\cite{LiXray}, etc.
(Some reviews are given in References\cite{FisherShen,HuXu}.)
The rotational symmetry breaking leads to a nematic phase transition
at critical temperature $T_c$.  This nematic phase transition shows close correlation
to the structural and magnetic phase transitions\cite{Fernandes2010,Fernandes2013,Bohmer2014}.
The origin of the nematic phase transition and its correlation with
superconductivity are now active topics in the field.

One complexity to study the electronic nematicity comes from the fact that
the rotational symmetry breaking manifests itself in different channels simultaneously.
Table \ref{tab1} shows the various nematic parameters in experiments.
From Landau's principle of symmetry breaking, the multiplicity of the nematic parameters
can be easily understood. Consider the symmetry breaking of Fe-site symmetry group from
$D_{2d}$ to $D_2$ in Fe-1111 and Fe-122 families. Any function $O_{\eta}$ which is
invariant under the operations of $D_2$ but not that of $D_{2d}$ can be
defined as a nematic parameter. It can be defined mathematically as
\begin{equation}
O_{\eta} = \sum_{\gamma} C_{\gamma} \Psi_{\gamma},
P_g [O_{\eta}] = O_{\eta} ,
\forall g\in D_{2},  \label{eqn1.1}
\end{equation}
where $\Psi_{\gamma}$ are basis of one irreducible non-identity representation
of $D_{2d}$ and $P_g$ is the corresponding operation of a group element $g$ of $D_2$.

\begin{table*}
\caption{ Examples of the various nematic parameters $O_{\eta}$ in different experimental techniques.
Whether the corresponding responses are static (S) or dynamical (D) and
whether they are related to rotational symmetry breaking in spatial (S) or magnetic (M)
spaces are also explicitly shown\cite{SuLi2013}.
$\sigma_{\alpha\alpha}$ are the conductivity along $\alpha$-axis, and
$\chi_{\alpha\alpha} (\mathbf{q},\omega) $ (or $\chi(\mathbf{q},\omega)$) are spin magnetic
susceptibility at momentum $\mathbf{q}$ and frequency $\omega$ with spin polarization along
$\alpha$-axis (without spin polarization). $\mathbf{Q}_1 = (\pi,0),
\mathbf{Q}_2=(0,\pi)$ and $\mathbf{Q}=\mathbf{Q}_1$ or $\mathbf{Q}_2$
are antiferromagnetic momenta.
$\omega_N$ in 1/T$_1$T is the nuclear resonance frequency which is too small
as compared to the electronic energy scales, $\gamma(\mathbf{q})$ is a form factor. } \label{tab1}
\begin{center}
\begin{ruledtabular}
\begin{tabular}{llcc}
Experimental technique &  Nematic parameter $O_{\eta}$ & S/D response
& 
\begin{tabular}{c}
Symmetry breaking \\
in S/M space
\end{tabular}  \\
\hline
Resistivity/Conductivity  &  $ \sigma_{xx} - \sigma_{yy}$
 & S/D  & S \\
ARPES &  $\Delta(\mathbf{k})\left\langle d^{\dag}_{\mathbf{k},xz} d_{\mathbf{k},xz}
 \pm d^{\dag}_{\mathbf{k},yz} d_{\mathbf{k},yz}\right\rangle$
 & S  & S \\
INS & & & \\
 \hspace{0.1cm} spin flip & $\chi_{xx}(\mathbf{Q},\omega) - \chi_{yy}(\mathbf{Q},\omega)$
  & D & M \\
  \hspace{0.1cm} spin wave & $\chi(\mathbf{q},\omega)$ in $\mathbf{q}$ variation & D & S \\
  \hspace{0.1cm} spin fluctuations & $\chi(\mathbf{Q}_1,\omega) - \chi(\mathbf{Q}_2,\omega)$
  & D & S \\
NMR & & & \\
\hspace{0.1cm} Knight shift & $ \chi_{xx} - \chi_{yy} $  & S & M \\
\hspace{0.1cm} 1/T$_1$T & $\alpha$-dependent $\sum_{\mathbf{q}} \gamma(\mathbf{q}) \chi_{\alpha\alpha}(\mathbf{q},\omega_N) $
 &  S & M \\
Torque &   $ \chi_{xx} - \chi_{yy} $ & S & M \\
Raman & $\left\langle \rho^{\mu} \rho^{\mu} \right\rangle$, $\rho^{\mu}$ Raman density
    & D & S  \\
STM/STS &  local density of states (LDOS) & S & S
\end{tabular}
\end{ruledtabular}
\end{center}
\end{table*}

The various nematic parameters show diverse manifestations of the
electronic nematicity in FeSCs. In ARPES, the nematicity shows
itself as orbital-relevant band shift\cite{ZXShen,YiShen2011,Feng2012,YiShen2014,DingH2015, LuDH2015}.
In inelastic neutron scattering (INS), it manifests anisotropic spin-wave
excitations (dispersion and damping)\cite{HarrigerDai2012}
and nematic dynamical spin fluctuations\cite{LuDaiScience2014,McQueeney2014},
the much low-energy part of the latter is also shown in 1/T$_1$T\cite{Fu2012}. In charge transport
resistivity and conductivity, the nematicity shows as a combined effect
of the anisotropic Fermi velocity and the anisotropic microscopic
scattering. The latter is proven as a dominant
factor for the nematicity in STM/STS\cite{AllanDavis,Rosenthal,Singh2013,AndersenPRL2014}.
It is also remarkable that the nematicity in optical conductivity shows in a very large
energy range from zero frequency to high energy of about $2$eV\cite{Uchida2011}.

The diverse experimental manifestations leads to hot debates on the primary driving mechanism
of the nematicity in FeSCs.
Among the various nematic parameters, the Ising spin (localized or itinerant) nematic order
\cite{HuXu,FangHu2008,CKXu,Fernandes2012,Fernandes2014}
and the orbital nematic order \cite{CCLee,Phillips,SuLi2014} are the most
popular candidates as the primary one. Another potential nematic order is related
to the Fermi-surface Pomeranchuk instability\cite{Zhai2009}.
Now it is still elusive which one is the primary driving force
and how these nematic orders correlate with each other.

In this manuscript, we will focus our study on the orbital nematic order.
From symmetry analysis we proposed recently a general form of the orbital nematic order
$O=\sum_{i a j b} F_{i a,j b} d^{\dag}_{i a} d_{j b} $, which involves a local
form factor $F_{i a,j b}$ ($i,j$ denote lattice site  and $a,b$ denote 3d orbital)\cite{SuLi2014}.
The local form factor $F_{i a,j b}$ can be on-site $s$-wave, nearest-neighbor $d$-wave
and nearest-neighbor extended $s^{\prime}$-wave, etc.
These orbital nematic orders can be defined in momentum space as,
\begin{eqnarray}
 O_{s} &=& \sum_{\mathbf{k}} \left( d^{\dag}_{\mathbf{k},xz} d_{\mathbf{k},xz}
 -d^{\dag}_{\mathbf{k},yz} d_{\mathbf{k},yz}\right) , \nonumber \\
 O_{d} &=& \sum_{\mathbf{k}} \Delta_d (\mathbf{k})\left( d^{\dag}_{\mathbf{k},xz} d_{\mathbf{k},xz}
 +d^{\dag}_{\mathbf{k},yz} d_{\mathbf{k},yz}\right) , \label{eqn1.4} \\
 O_{s^{\prime}} &=& \sum_{\mathbf{k}} \Delta_{s^{\prime}}(\mathbf{k})\left( d^{\dag}_{\mathbf{k},xz} d_{\mathbf{k},xz}
 -d^{\dag}_{\mathbf{k},yz} d_{\mathbf{k},yz}\right) , \nonumber
\end{eqnarray}
where $\Delta_d(\mathbf{k}) = (\cos k_x - \cos k_y)/2$ and
$\Delta_{s^{\prime}}(\mathbf{k}) = (\cos k_x + \cos k_y)/2$.

The on-site orbital nematic order $O_s$ has been taken extensively as a nematic order in FeSCs.
However a detailed investigation on the ARPES data shows that the orbital-dependent
band shift is strong momentum-dependent\cite{ZXShen,YiShen2011,Feng2012,YiShen2014,DingH2015, LuDH2015}.
In BaFe$_2$As$_2$\cite{ZXShen,YiShen2014} and FeSe\cite{DingH2015, LuDH2015} families,
the band shift is much larger at momentum near $\mathbf{Q}_1=(\pi,0)$
and $\mathbf{Q}_2=(0,\pi)$ than that near $\Gamma$ point $\mathbf{k}=(0,0)$.
Obviously it can not be accounted for by a simple on-site orbital nematic order $O_s$.
This unusual momentum-dependent band shift can be naturally interpreted by
a {\it bond} $d$-wave orbital nematic order $O_d$ as we have proposed\cite{SuLi2014,LiangHu2015,HuWang2015}.
The bond orbital nematic orders we have introduced for the nematic state of FeSCs
are very similar to the bond charge-density-wave (CDW) order which is
proposed for the CDW state of the cuprate superconductors\cite{SachdevCDW}.
Compared to the on-site order, the {\it bond} order can avoid the energy enhancement
from strong local Coulomb interaction, thus makes the system in a more stable and
lower energy state. A similar idea has been proposed in a recent study\cite{HuWang2015}.

In our recent study, we have shown that the orbital nematic order can
enhance the condensation energy of the magnetic state\cite{SuLi2014}.
In this manuscript, we will follow this study to investigate the
influence of the orbital nematic order on the nematic spin correlations.
Recent INS experiment shows strong nematic spin fluctuations
in FeSCs, where the spin fluctuations at antiferromagnetic (AFM) momentum
$\mathbf{Q}_1$ are much larger than those at AFM momentum $\mathbf{Q}_2$\cite{LuDaiScience2014}.
Moreover the integrated strength of the AFM spin fluctuations is enhanced sharply
when across the nematic phase transition\cite{McQueeney2014}.
In this manuscript, we introduce a finite orbital nematic order for the nematic state
of FeSCs. We show that a finite orbital nematic order can drive a significant spin correlation 
nematicity and can enhance the integrated intensity of the spin fluctuations as observed 
in the INS experiments\cite{LuDaiScience2014,McQueeney2014}.
Our results show that the orbital nematic order has strong effect on the spin
nematicity, and it should not be taken as a simple secondary effect caused
by the symmetry breaking from the spin nematicity.

Our manuscript is arranged as following.
In Sec. \ref{sec2} we specify a simplified model Hamiltonian for the nematic state of FeSCs.
In Sec. \ref{sec3} we study the spin correlations with a finite orbital
nematic order. Sec. \ref{sec5} shows our discussion and conclusion.

\section{Model Hamiltonian} \label{sec2}

At present a well-defined theory for the nematic state of FeSCs is absent.
This is due to the entanglement of the various nematic responses as observed 
in diverse experimental probes.
In our study, we will simplify our focus on how the orbital nematic order
makes influence on the spin fluctuations in FeSCs.
We thus introduce a simplified Hamiltonian which includes a mean-field
part $H_0$ for the ordered nematic state and an interacting part $H_I$ to
account for the interaction-renormalized spin fluctuations.

The mean-field Hamiltonian $H_0$ is defined as
$H_0 = H_t + H_{mf}$, where $H_t$ describes the electronic band
structure with the $3d$ orbitals involved and $H_{mf}$ is the mean-field
contribution from a finite orbital nematic order.
Following Kuroki {\it et al.}\cite{Kuroki}, we define $H_t$ as
\begin{equation}
H_t = \sum_{iajb \sigma} t_{ia,jb} d^{\dag}_{ia \mu} d_{jb \mu} , \label{eqn2.1}
\end{equation}
where $d_{ia \mu}$ and $d^{\dag}_{ia \mu}$ are the annihilation and creation operators
respectively for $3d$ electrons. The subscripts $i/j, a/b$ and $\mu$ are indices for
lattice site, orbital and spin degrees of freedom, respectively.
$H_{mf}$ describes the orbital nematic state in mean-field approximation
and is defined by
\begin{equation}
H_{mf} = -\Delta \ \ O_{\eta} , \label{eqn2.2}
\end{equation}
where $O_{\eta}$ is given in Eq. (\ref{eqn1.4}).
Here we make approximation that an unknown microscopic origin of the orbital nematic order
from such as multi-orbital Hubbard interaction or beyond is not considered.
The recent ARPES data\cite{DingH2015, LuDH2015} shows that the band shift
has a mean-field-like temperature dependence, we thus propose that the orbital nematic
order follows $\Delta = \Delta_{0}\left( 1-T/T_c \right)^{1/2}$.
In our study, we set $100$meV as an energy unit and set $\Delta_0 = 0.33$ ($\sim 33$meV) following
ARPES data of Yi {\it et al.}\cite{ZXShen} and $T_c=0.13$ ($\sim 140$K).
Chemical potential is set as $\mu_{F} = 109$ ($\sim 10.9$eV) for $3d$ electron number 
per-site $n_d \simeq 6.02$.
The constants $k_B= \hbar = 1$.

The interacting Hamiltonian $H_I$ is defined as a multi-orbital Hubbard interaction,
\begin{eqnarray}
&&H_I= U\sum_{ia} n_{ia\uparrow} n_{ia\downarrow}
+ \left(U^{\prime}-\frac{J}{2}\right) \sum_{i,a<b}n_{ia}n_{ib} \label{eqn2.3}  \\
&&- 2 J \sum_{i,a<b}\mathbf{S}_{ia}\cdot\mathbf{S}_{ib}
+ J^{\prime}
\sum_{i,a<b} \left( d^{\dag}_{ia\uparrow}d^{\dag}_{ib\downarrow}
d_{ib\downarrow}d_{ia\uparrow} + h.c. \right).   \nonumber
\end{eqnarray}
It should be noted that the physical roles of $H_I$ in the renormalization
of the electronic band structure and in the formation of the orbital nematic order
have been assumed to be included in $H_t$ and $H_{mf}$, respectively.
In our following study, we will only consider the role of $H_I$ in
the spin correlations.

\section{Dynamical spin susceptibility} \label{sec3}

In this section, we will study the influence of the orbital nematic
order on the spin correlations in the nematic state without long-range
magnetic order in FeSCs. We will be interested in whether a finite
orbital nematic order can induce the spin nematic responses as
observed in the INS experiments\cite{LuDaiScience2014,McQueeney2014}.

Introducing a generalized multi-orbital spin operator
$$
\mathbf{S}_{a_1 a_2} = \frac{1}{\sqrt{N}} \sum_{\mathbf{k} \mu_1 \mu_2} d^{\dag}_{\mathbf{k}+\mathbf{q} a_1 \mu_1}
\left( \frac{ \boldsymbol{\sigma} }{2}\right)_{\mu_1 \mu_2} d_{\mathbf{k}a_2 \mu_2}
$$
with $\boldsymbol{\sigma}$ being the Pauli matrix,
we define the spin (transverse) susceptibility as
\begin{equation}
\chi^{(+-)}_{a_1 a_2,a_3 a_4} (\mathbf{q},\tau) = \langle T_{\tau} S^{+}_{a_1 a_2}(-\mathbf{q},\tau)
S^{-}_{a_4 a_3}(\mathbf{q},0) \rangle .  \label{eqn3.1}
\end{equation}
Since the scattering cross section of INS off FeSCs is determined by
the dynamical spin correlation function $\langle S^{\alpha}_{a_1 a_1}(-\mathbf{q},t)
S^{\alpha}_{a_2 a_2}(\mathbf{q},0) \rangle$ with $\alpha = x, y, z$,
the neutron spin responses in the nematic state without long-range magnetic order
is determined by
\begin{equation}
\chi(q) = \frac{1}{2} \sum_{a_1 a_2} \chi^{(+-)}_{a_1 a_1 ; a_2 a_2} (q) ,   \label{eqn3.2}
\end{equation}
where $q = (\mathbf{q}, i\nu_n)$.
Because the nematic state of FeSCs is near magnetic instability,
we should consider the additional random phase approximation (RPA) effect from the
multi-orbital Hubbard interaction $H_I$.
We thus have\cite{YadaKontani}
\begin{equation}
\chi_{RPA}^{(+-)} (q) = \left(1 - \chi^{(+-)}_{0} (q) \hat{V} \right)^{-1} \chi^{(+-)}_{0}(q) ,
\label{eqn3.3}
\end{equation}
where $\chi^{(+-)}_{0} (q)$ is the bare spin susceptibility corresponding
to the mean-field Hamiltonian $H_0$, and the interaction matrix $\hat{V}$ is defined by
\begin{eqnarray}
\hat{V}_{a_1 a_2 , a_3 a_4} = \left\{
\begin{array} {c c}
U, &   {\rm if} \ \ a_1 = a_2 = a_3 = a_4 , \\
U^{\prime}, &   {\rm if} \ \ a_1 = a_3 \not= a_2 = a_4 , \\
J, &   {\rm if} \ \ a_1 = a_2 \not= a_3 = a_4 , \\
J^{\prime}, &   {\rm if} \ \ a_1 = a_4 \not= a_2 = a_3 .
\end{array}
\right.  \nonumber
\end{eqnarray}
It should be noted that the spin susceptibility thus obtained includes the influence
of the orbital nematic order, but has no feedback effect.
If the origin of the orbital nematic order is spin irrelevant, this formalism is
well defined. But if it is spin relevant, both as a secondary effect
or strong interplay with the spin nematicity, this formalism
is approximate at zero-th order. In the latter case, the contribution from the spin
interaction to the orbital nematic order has been assumed to be included in the mean-field
Hamiltonian $H_{mf}$. Only higher-order spin fluctuations are ignored in the
feedback effect to the orbital nematic order.

The calculation of the bare spin susceptibility $\chi^{(+-)}_{0}(q)$ is straight forward,
which can be easily obtained as
\begin{equation}
\chi^{(+-)}_{0; a_1 a_2, a_3 a_4} (q) = \frac{1}{N}\sum_{\mathbf{k} m n \mu_1 \mu_2}
C^{mn}_{\mu_1 \mu_2}(\mathbf{k},\mathbf{q})
F^{m n}_{\mu_1 \mu_2}(i \nu_n, \mathbf{k}, \mathbf{q}), \label{eqn3.4}
\end{equation}
with
\begin{eqnarray}
C^{mn}_{\mu_1 \mu_2}(\mathbf{k},\mathbf{q})
&&= U^{\dag}_{n a_1} (\mathbf{k} \mu_1) \frac{\boldsymbol{\sigma}^{+}_{\mu_1 \mu_2}}{2}
U_{a_2 m} (\mathbf{k+q},\mu_2)  \nonumber \\
&& \cdot U^{\dag}_{m a_4} (\mathbf{k+q},\mu_2) \frac{\boldsymbol{\sigma}^{-}_{\mu_2 \mu_1}}{2}
U_{a_3 n} (\mathbf{k} \mu_1)  \label{eqn3.5}
\end{eqnarray}
and
\begin{equation}
F^{m n}_{\mu_1 \mu_2} (i \nu_n, \mathbf{k}, \mathbf{q})
 = \frac{f(E_{\mathbf{k+q}, m \mu_2}) - f(E_{\mathbf{q}, n \mu_1})}
 {i \nu_n - E_{\mathbf{k+q}, m \mu_2} + E_{\mathbf{k}, n \mu_1}} .\label{eqn3.6}
\end{equation}
Here $E_{\mathbf{k}, m \mu}$ is the m-th eigenvalue of $H_0$ and $U(\mathbf{k} \mu)$
is the transformation matrix to diagonalize $H_0$.
$f(x)$ is the Fermi-distribution function.
In our study, the electronic band structure of $H_t$ is defined as
Kuroki {\it et al.}\cite{Kuroki}. The parameters in the multi-orbital Hubbard
interaction $H_I$ are chosen not far away from a magnetic instability,
and according to Kuroki {\it et al.}\cite{Kuroki} we set
$U=9.0$ ($0.9$eV), $U^{\prime} = 6.6$ ($0.66$eV), $J=J^{\prime}=1.2$ ($0.12$eV).
The rotation of the imaginary frequency to the real one is defined by
$i \nu_n \rightarrow \omega + i\delta_\tau$ with $\delta_\tau=0.05$.

\begin{figure}[hbp]
\begin{center}
\includegraphics [angle=0,width=1.0\columnwidth,clip=true]{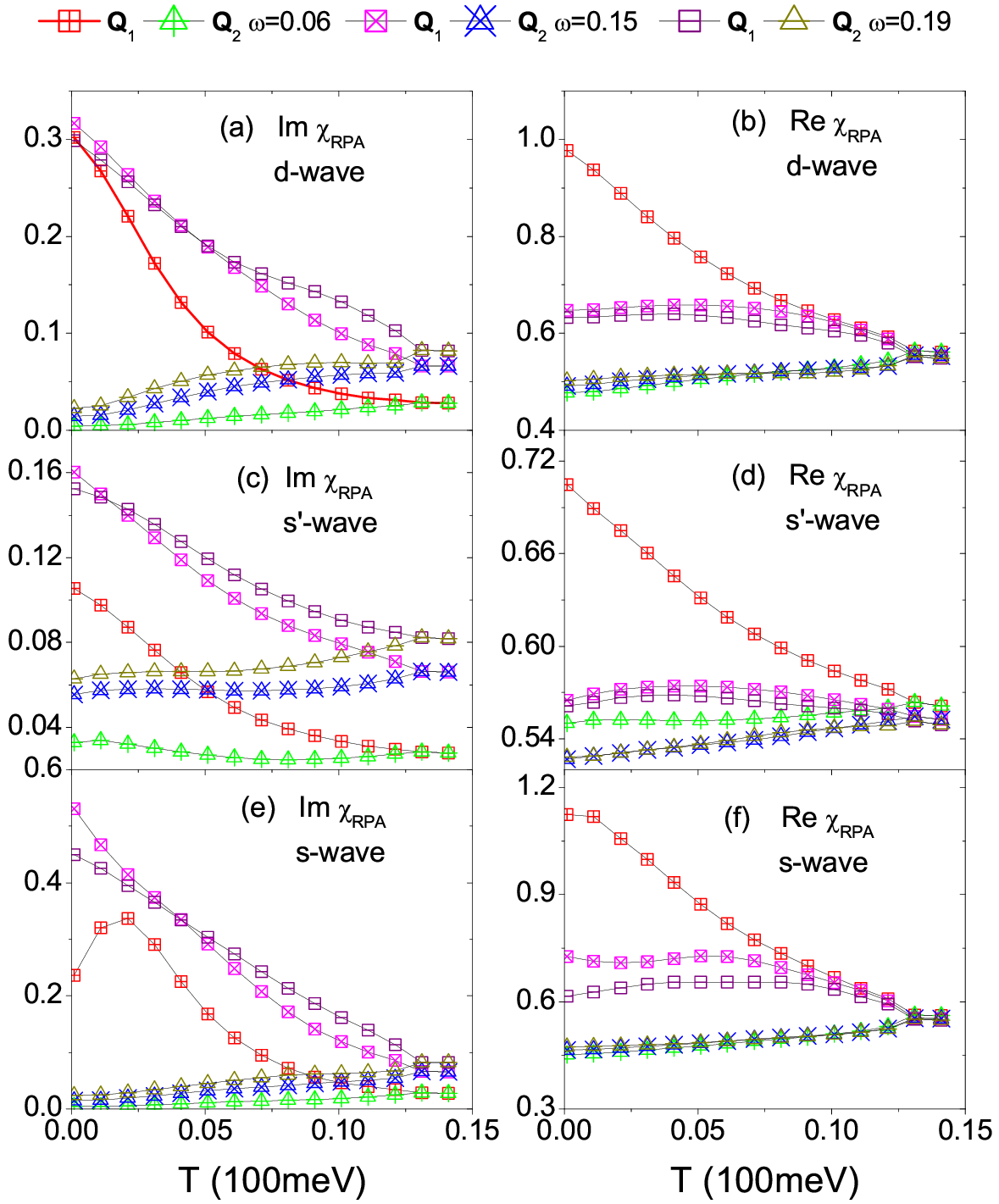}
\caption{ (Color online) Nematic spin responses shown at
AFM momenta $\mathbf{Q}_{1}=(\pi,0)$ and $\mathbf{Q}_2=(0,\pi)$.
(a) imaginary part and (b) real part of the spin susceptibility $\chi_{RPA}(\mathbf{q},\omega)$
with $d$-wave orbital order, (c) and (d) for $s^{\prime}$-wave and (e) and (f) for
$s$-wave.  The critical temperature of the nematic phase transition is $T_c=0.13$ and
the orbital nematic order at $T=0$K is $\Delta_0 =0.33$. In our study, we set $100$meV
as an energy unit. } \label{fig3.1}
\end{center}
\end{figure}

\begin{figure}[htbp]
\begin{center}
\includegraphics [angle=0,width=1.0\columnwidth,clip=true]{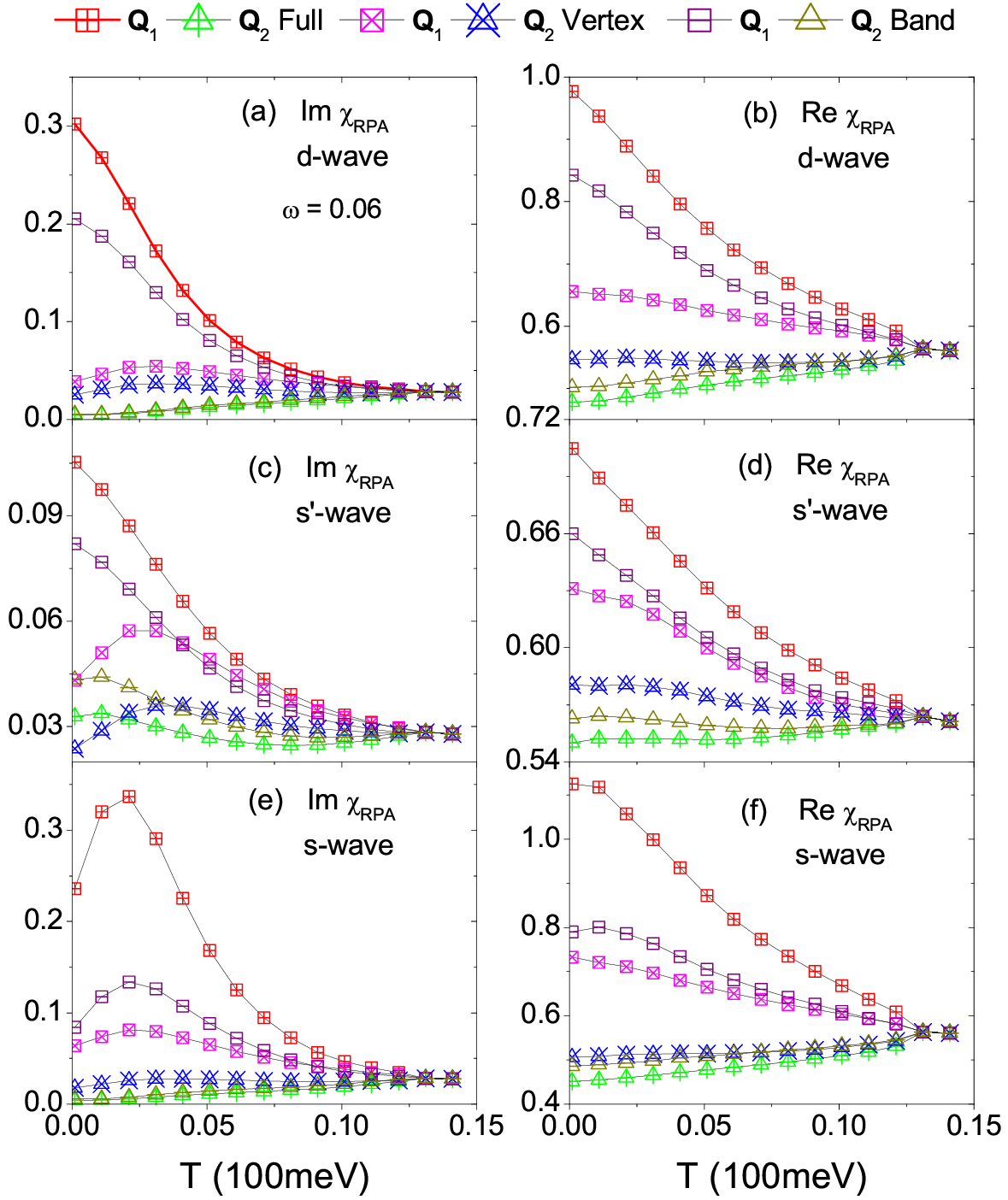}
\caption{ (Color online)   The {\it vertex } and {\it band} effects of
the orbital nematic order on the spin responses. Also shown includes the
{\it full} effect of the orbital nematic order. Frequency is fixed at $\omega = 0.06$
and other parameters are  same to Fig. \ref{fig3.1}.  } \label{fig3.2}
\end{center}
\end{figure}

In Fig. \ref{fig3.1}, we plot the temperature-dependent spin susceptibility
at AFM $\mathbf{Q}_1$ and $\mathbf{Q}_2$ with different orbital nematic orders.
For comparison with the INS experiment\cite{LuDaiScience2014}, three frequencies
$\omega = 6$meV, $15$meV and $19$meV are considered.
It is obvious that in the nematic state with a finite orbital nematic order,
the spin fluctuations break the tetragonal symmetry.
While the spin susceptibility at $\mathbf{Q}_1$ increases with deceasing temperature,
the spin susceptibility at $\mathbf{Q}_2$ decreases with temperature,
which is consistent to the INS experiment\cite{LuDaiScience2014}.
This result shows that a finite orbital nematic order can easily
lead to nematic spin fluctuations.

In the multi-orbital FeSCs, both the orbital character of the eigenstates and the band
dispersion have important influence on the dynamical spin responses. Thus the
orbital nematic order, which lifts the degeneracy of the $d_{xz}$ and $d_{yz}$ states,
would make influence on the nematic spin fluctuations from two
factors, the orbital character of the eigenstates and the band dispersion.
To distinguish these two factors, we consider the following two cases.
In the first case we call it the {\it vertex} effect, we set the orbital nematic
order zero in $F^{m n}_{\mu_1 \mu_2}$. In this case the orbital nematic order
plays role through the $U$ matrix which carries the orbital character of the
eigenstates. In the other case we call it the {\it band} effect, we set the orbital
nematic order zero in $C^{m n}_{\mu_1 \mu_2}$. In this case the role of
the orbital nematic order is mainly to modify the Fermi-surface nesting condition through
the band dispersion in function $F^{m n}_{\mu_1 \mu_2}$.
Fig. \ref{fig3.2} shows the  vertex and  band effects of the orbital nematic order on
the spin susceptibility. For comparison the full effect is also shown.
With a $d$-wave orbital nematic order, the dynamical spin nematicity is dominant by
the band effect through a modification of Fermi-surface nesting condition.
With a $s^{\prime}$- or $s$-wave nematic order, the vertex and band effects have
nearly same contributions in the spin response nematicity.
It should be noted that the vertex and band contributions of the orbital nematic 
order in the spin response nematicity are not additive.

\begin{figure}[htp]
\begin{center}
\includegraphics [angle=0,width=0.85\columnwidth,clip=true]{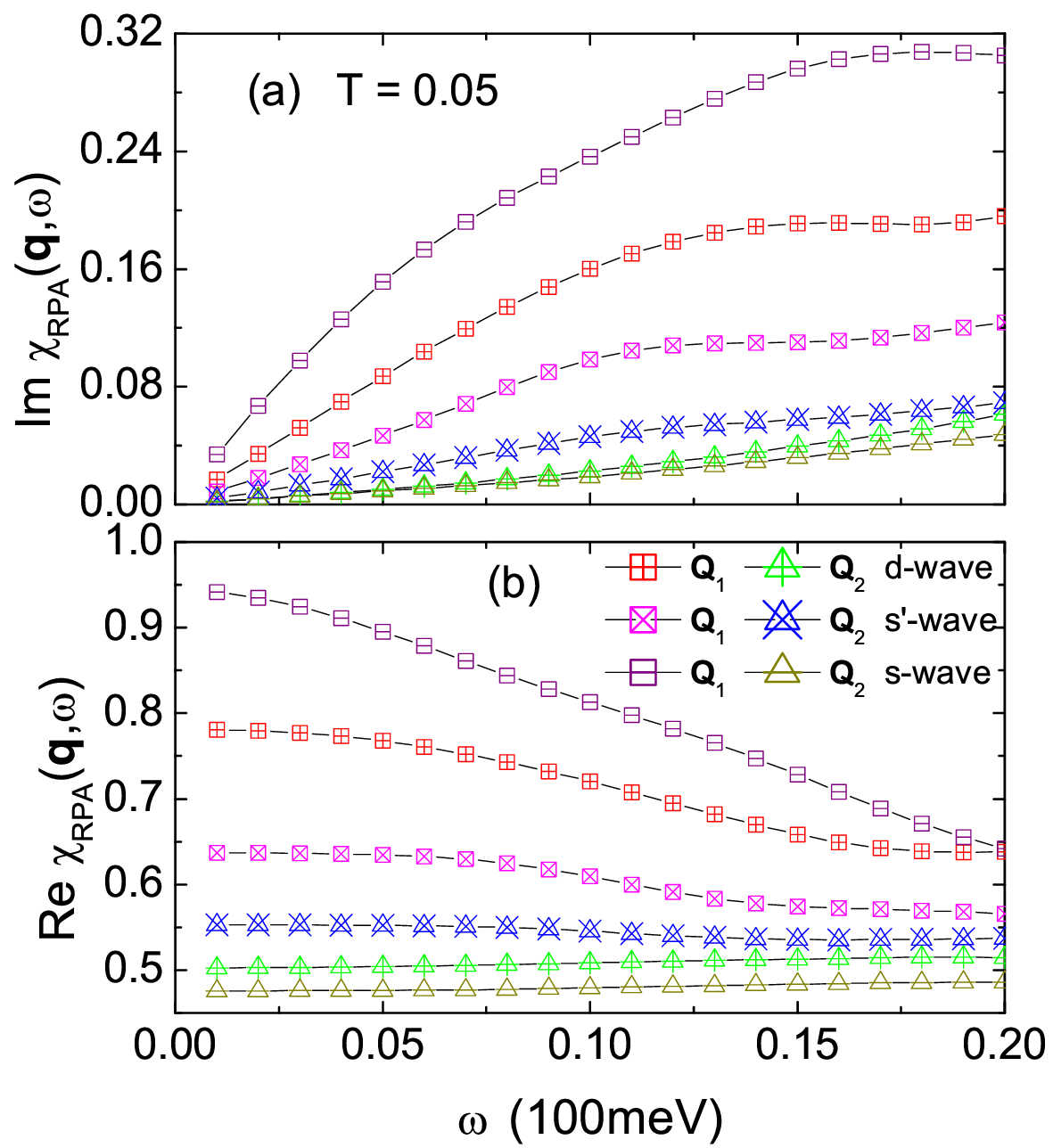}
\caption{(Color online)  Frequency-dependent spin susceptibility
at AFM momenta $\mathbf{Q}_{1}$ and $\mathbf{Q}_2$. Temperature is fixed
at $T=0.05$. Other parameters are same to Fig. \ref{fig3.1}.}
\label{fig3.3}
\end{center}
\end{figure}

In Fig. \ref{fig3.1} and Fig. \ref{fig3.2}, we also plot the real part of the
spin susceptibility. The orbital nematic order leads to a similar nematic behavior
of the real part of the spin response function.
In Fig. \ref{fig3.3}, we show the frequency dependence of the spin susceptibility.
In the range $\omega\in \left( 0, 0.2\right)$, the nematicity in the imaginary part
of the spin susceptibility increases with frequency, while that in the real part
decreases. At each frequency the $s^{\prime}$-wave orbital nematic order has a
weaker effect on the spin fluctuation nematicity than the $d$- and $s$-wave
orbital nematic orders.

\begin{figure*}[htp]
\begin{center}
\includegraphics [angle=0,width=1.8\columnwidth,clip=true]{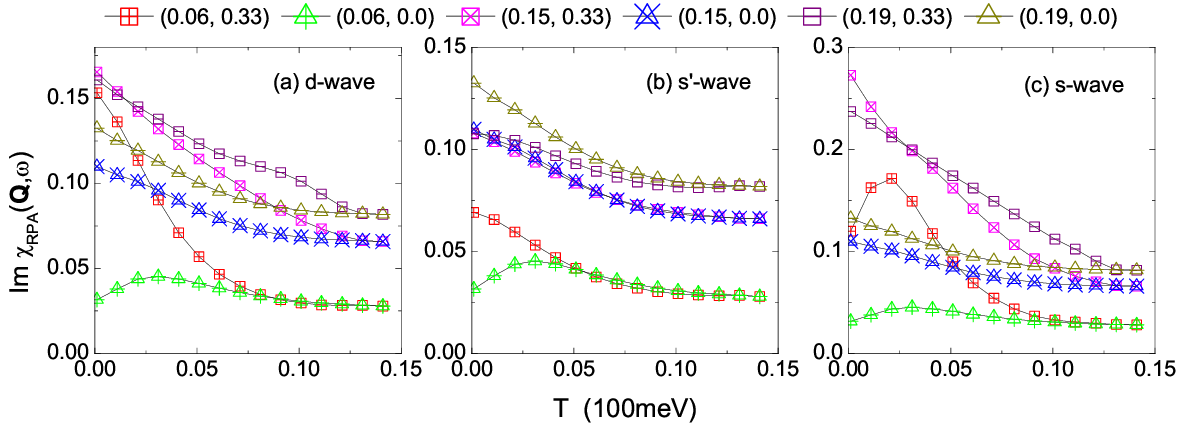}
\caption{(Color online)  Comparison of the imaginary part of the AFM spin susceptibility
with and without orbital nematic order. ${\rm Im} \chi_{RPA}(\mathbf{Q},\omega)
=\left[{\rm Im} \chi_{RPA}(\mathbf{Q}_1,\omega)+{\rm Im} \chi_{RPA}(\mathbf{Q}_2,\omega)\right]/2 $.
The data with different $\omega$ and $\Delta_0$ are shown in form  $(\omega, \Delta_0)$. }
\label{fig3.4}
\end{center}
\end{figure*}

Recently neutron scattering measurement shows strong effect of the electronic
nematicity on the spin fluctuations in {\it twinned} FeSCs\cite{McQueeney2014}.
When temperature decreases below the nematic critical temperature $T_c$ and
the system enters into the nematic state, the dynamical spin responses at AFM momentum $\mathbf{Q}_{AF}$
show a strong increase and exhibit a maximum at lower AFM transition temperature.
It shows an enhancement effect of the electronic nematicity on the dynamical spin fluctuations.
To understand this enhancement effect, we calculate the AFM spin susceptibility defined as
$$
\chi_{RPA}(\mathbf{Q},\omega)
=\frac{1}{2}\left[ \chi_{RPA}(\mathbf{Q}_1,\omega)+ \chi_{RPA}(\mathbf{Q}_2,\omega)\right].
$$
Note that in the AFM spin susceptibility for twinned FeSCs, the AFM momenta $\mathbf{Q}_1$ and $\mathbf{Q}_2$
can not be distinguished. Fig. \ref{fig3.4} shows our calculation.
Clearly the $d$- or $s$-wave orbital nematic order can enhance the AFM
spin fluctuations, which is consistent to the experiment observation\cite{McQueeney2014}.
As a comparison, the $s^{\prime}$-wave orbital nematic order has a
much weak effect on the integrated AFM spin fluctuations.
This result implies that the $s^{\prime}$-wave orbital nematic order may not be
a dominant orbital order in LaFeAsO and Ba(Fe$_{1-x}$Co$_x$)$_2$As$_2$ families.

\section{Discussion and conclusion} \label{sec5}

In this manuscript, we have studied the influence of the orbital nematic order
on the spin responses. It shows that orbital nematic order
can readily lead to the spin response nematicity and can
enhance the strength of the AFM spin fluctuations. Our results are
consistent to the INS experiments\cite{LuDaiScience2014,McQueeney2014}
and the recent theoretical studies\cite{Andersen2014,Mukherjee2015,Kreisel2015}.

Although the orbital nematic order is proven to has important influence on
the spin correlations, it should not be taken as proof that the orbital
nematic order is the primary driving force for the electronic nematicity in FeSCs.
This is because that the spin nematicity itself can easily occur without orbital
degree of freedom involved\cite{HuXu,FangHu2008,CKXu,Fernandes2012,Fernandes2014}.
Moreover, the microscopic mechanism of the orbital nematic order is still elusive.
Since the nematic state is proximate to and can coexist with the magnetic state
in FeSCs, the orbital nematic order may be spin-interaction relevant. In this case,
the orbital and spin nematic orders would be in strong entanglement.

Whether or not the orbital nematic order is a primary driving force for the
electronic nematicity in FeSCs, our study shows that the orbital nematic order
has strong effect on the nematicity and the integrated strength of
the spin fluctuations. Moreover, our recent study shows that
the orbital nematic order can enhance significantly the condensation energy
of the magnetic state\cite{SuLi2014}. We thus conclude that the orbital nematic
order has important influence on the spin correlations in FeSCs and
should not be taken as an unimportant secondary effect of the spin nematicity.
This conclusion also indicates that the orbital nematic order may play
important roles in the magnetism and the superconductivity in FeSCs.

This work was supported by the National Natural Science Foundation of China
(Grant Nos. 10974167, 11304269, 10774187, 11034012) and
National Basic Research Program of China (Grant No. 2010CB923004).


\end{document}